\documentclass[a4paper,twoside,twocolumn,english,aps,superscriptaddress,showpacs,prl]{revtex4}
\usepackage[T1]{fontenc}
\usepackage[latin1]{inputenc}
\usepackage{color}
\usepackage{graphicx}
\usepackage{amssymb}

\makeatletter

\providecommand{\boldsymbol}[1]{\mbox{\boldmath $#1$}}

\newcommand{\lyxdot}{.}

\usepackage{babel}
\makeatother
\begin{document}

\title{Few-boson dynamics in double wells: From single-atom to correlated-pair
tunneling}

\author{Sascha Zöllner}

\email{sascha.zoellner@pci.uni-heidelberg.de}

\affiliation{Theoretische Chemie, Universit\"{a}t Heidelberg, Im Neuenheimer
Feld 229, 69120 Heidelberg, Germany}

\author{Hans-Dieter Meyer}

\affiliation{Theoretische Chemie, Universit\"{a}t Heidelberg, Im Neuenheimer
Feld 229, 69120 Heidelberg, Germany}

\author{Peter Schmelcher}

\affiliation{Theoretische Chemie, Universit\"{a}t Heidelberg, Im Neuenheimer
Feld 229, 69120 Heidelberg, Germany}

\affiliation{Physikalisches Institut, Universit\"{a}t Heidelberg, Philosophenweg
12, 69120 Heidelberg, Germany}

\begin{abstract}
We investigate few-boson tunneling in a one-dimensional double well,
covering the full crossover from weak interactions to the fermionization
limit of strong correlations. Based on exact quantum-dynamical calculations,
it is found that the tunneling dynamics of two atoms evolves from
Rabi oscillations to correlated pair tunneling as we increase the
interaction strength. Near the fermionization limit, fragmented-pair
tunneling is observed and analyzed in terms of the population imbalance
and two-body correlations. For more atoms, the tunneling dynamics
near fermionization is shown to be sensitive to both atom number and
initial configuration.
\end{abstract}

\date{November 22, 2007}


\maketitle

The double well is a paradigm model for some of the most fundamental
quantum effects, like interference or tunneling. Using ultracold atoms,
it has become possible to study this system at an unprecedented level
of control. This has lead, e.g., to the observation of Josephson oscillations
\cite{albiez05,milburn97,smerzi95} and nonlinear self-trapping \cite{albiez05,anker05,javanainen86}
of Bose-Einstein condensates. In the first case, the weakly interacting
atoms---prepared mostly in one well---simply tunnel back and forth
between the two wells in analogy to a Josephson current. However,
above a critical interaction strength, the atoms essentially remain
trapped in that well for the experimental lifetime even though they
repel each other. On the few-body level, this resembles the situation
of repulsive atom pairs, whose stability \cite{winkler06} and dynamics
\cite{foelling07} have recently been observed.

All of these effects are confined to the regime of relatively weak
interactions, where the dynamics can be understood qualitatively (up
to phases) by means of a single parameter: the number of atoms in
one well. However, interactions in ultracold atoms can be adjusted
experimentally over a wide range, e.g., via Feshbach resonances \cite{koehler06}.
In particular, in one dimension (1D) one can tune the effective interaction
strength at will by exploiting a confinement-induced resonance \cite{Olshanii1998a},
which makes it possible to explore the limit of strong correlations.
If the bosons repel each other infinitely strongly, they can be mapped
to noninteracting fermions \cite{girardeau60} in the sense that the
exclusion principle mimics the \emph{hard-core} interaction. While
local properties like the densities are shared with their fermionic
counterparts, nonlocal aspects such as their momentum distribution
are very different. Sparked also by its experimental demonstration
\cite{paredes04,kinoshita04}, this \emph{fermionization} has attracted
broad interest (see \cite{girardeau00,zoellner06a} and Refs. therein). 

In this Letter, we investigate the case where a few atoms are loaded
into the same well and explore the tunneling dynamics as we vary the
interaction strength from zero up to the fermionization limit. For
two atoms, we show that the character of the tunneling changes from
Rabi oscillations to correlated pair tunneling. Near fermionization,
the strongly interacting atoms tunnel back and forth as a fragmented
pair. For three or more atoms, the tunneling dynamics turns out to
depend strongly on the atom number and the initial imbalance.

\paragraph{Model and computational method}

The double-well dynamics is described by the many-body Hamiltonian
$H=\sum_{i=1}^{N}\left[\frac{1}{2}p_{i}^{2}+U(x_{i})\right]+g\sum_{i<j}\delta_{\sigma}(x_{i}-x_{j})$.
Here the double well $U(x)=\frac{1}{2}x^{2}+h\delta_{w}(x)$ is modeled
as a superposition of a harmonic oscillator and a central barrier
shaped as a Gaussian $\delta_{w}(x)=e^{-x^{2}/2w^{2}}/\sqrt{2\pi}w$
(we choose $w=0.5$ and $h=8$, where harmonic-oscillator units are
employed throughout.) The effective interaction resembles a 1D contact
potential \cite{Olshanii1998a}, but is mollified with a Gaussian
$\delta_{\sigma=0.05}$  so as to alleviate the well-known numerical
difficulties of the $\delta$ function. We focus on repulsive forces
$g\in[0,\infty)$.

\paragraph*{}

Our goal is to investigate the few-atom quantum dynamics in the crossover
to the highly correlated fermionization limit $g\to\infty$ in a numerically
\emph{exact} fashion. This is a challenging task, and most studies
on the double-well dynamics so far have relied on two-mode models
\cite{milburn97,tonel05} valid for sufficiently weak coupling. Our
approach rests on the Multi-Configuration Time-Dependent Hartree method
\cite{mey90:73}, a wave-packet dynamics tool which has been applied
successfully to few-boson systems (see \cite{zoellner06a} for details).

\paragraph*{From uncorrelated to pair tunneling}

\begin{figure}
\includegraphics[width=0.8\columnwidth,keepaspectratio]{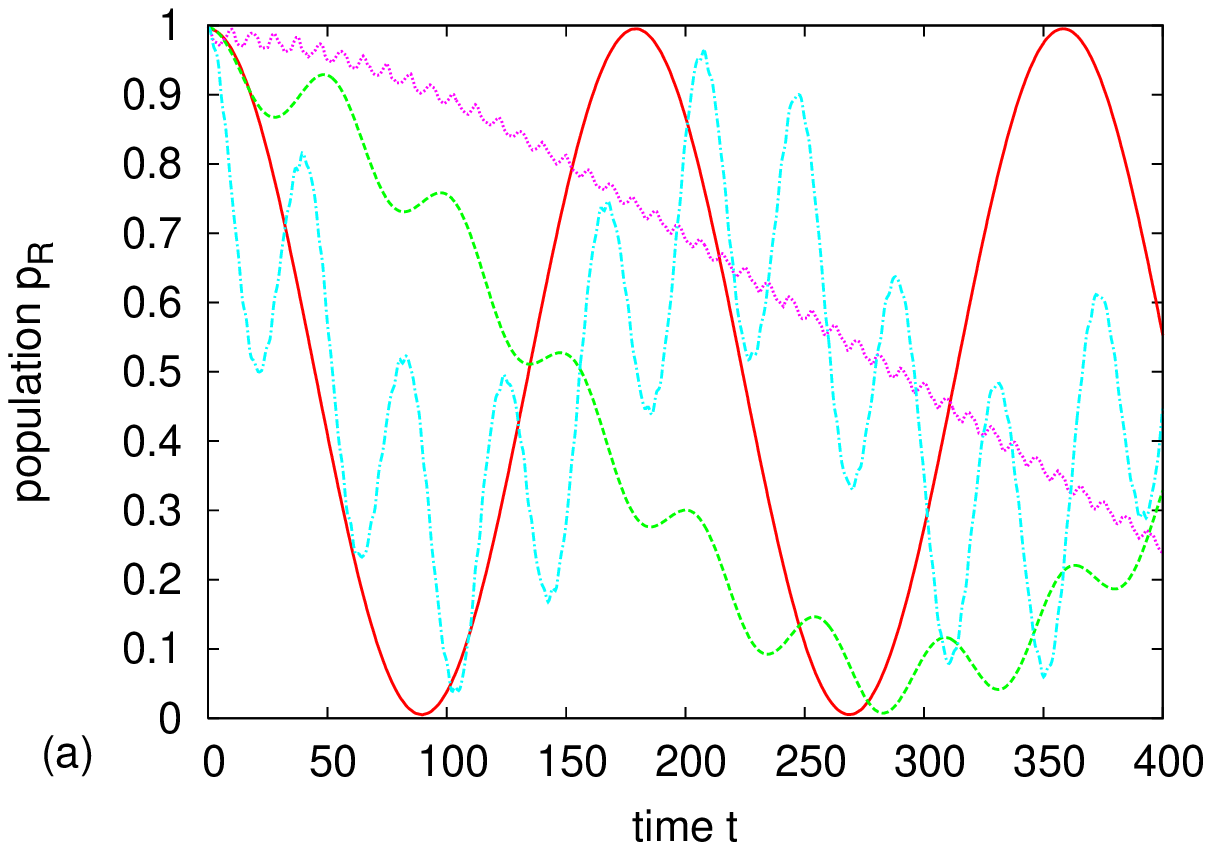}

\includegraphics[width=0.8\columnwidth,keepaspectratio]{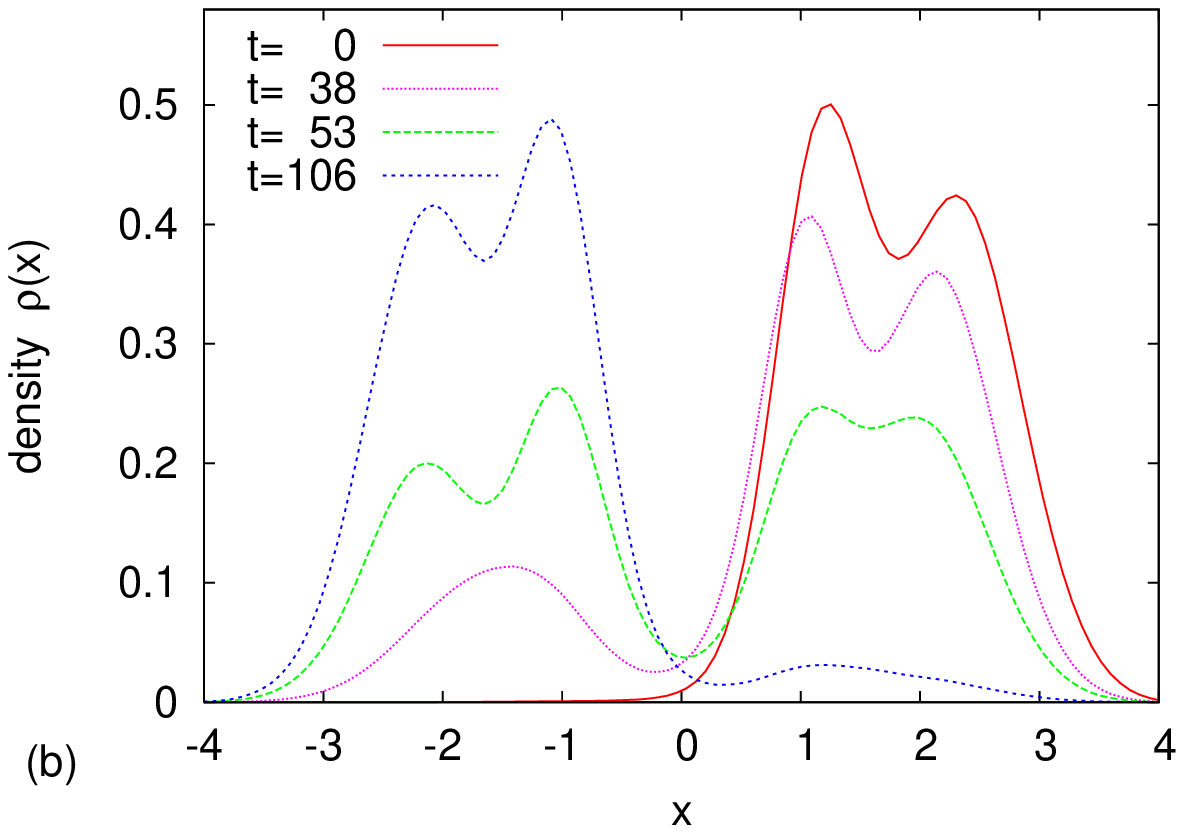}

\caption{(color online) Two-atom dynamics. (a) Relative population of the
right-hand well over time, $p_{\mathrm{R}}(t)$, for different interaction
strengths $g=0$ (\textcolor{red}{---}), $g=0.2$ (\textcolor{green}{-
- -}), $g=4.7$ (\textcolor{magenta}{${\color{magenta}\cdots}$}),
and $g=25$ (${\color{cyan}-\cdot-}$). (b) Snapshots of the one-body
density $\rho(x)$ for different times $t$ in the fermionized case
$g=25$. (\emph{All quantities in harmonic-oscillator units throughout,
see text.}) \label{cap:pop-N2}}
\end{figure}

To prepare the initial state $\Psi(0)$ with a population imbalance---in
our case, such that almost all atoms reside in the right-hand well---we
make that side energetically favorable by adding a linear external
potential $-d\cdot x$ ($d>0$) and let the system relax to its ground
state $\Psi_{0}^{(d>0)}$. For sufficiently large $d$, this amounts
to preparing nearly all atoms in one well. To study their time evolution
in the \emph{symmetric} double well, in our simulations the asymmetry
will be ramped down, $d\to0$, within some time $\tau>0$. 

Let us now study how the tunneling changes as we pass from uncorrelated
tunneling ($g=0$) to tunneling in the presence of correlations and
finally to the fermionization limit ($g\to\infty$). It is natural
to first look at the conceptually clearest situation where $N=2$
atoms initially reside in the right-hand well. Absent any interactions,
the atoms simply \emph{Rabi-}oscillate back and forth between both
wells, which materializes in the percentage of atoms in the right
well $p_{\mathrm{R}}(t)=\langle\Theta(x)\rangle_{\Psi(t)}=\int_{0}^{\infty}\rho(x;t)dx$
($\rho$ being the one-body density) or, correspondingly, the population
imbalance $\delta=p_{\mathrm{R}}-p_{\mathrm{L}}=2p_{\mathrm{R}}-1$.
By contrast, if the atoms repel each other, then the tunneling process
will be modified, as can be seen in Fig.~\ref{cap:pop-N2}(a). For
$g=0.2$, one sees that the tunneling oscillations have become a two-mode
process: There is a fast (small-amplitude) oscillation which modulates
a much slower oscillation in which the atoms eventually tunnel completely
($p_{\mathrm{R}}\approx0$). In case $g$ is increased further, we
have found that the tunnel period becomes indeed so long that complete
tunneling is hard to observe. E.g., at $g=1.3$ the period is as large
as $2\times10^{3}$. What remains is a very fast oscillation with
only a minute amplitude -- the two-body analogue of quantum self-trapping.
As we go over to much stronger couplings (see $g=4.7$), we find that
the time evolution becomes more and more complex, even though this
is barely captured in the reduced quantity $p_{\mathrm{R}}$ {[}Fig.~\ref{cap:pop-N2}(a)].
What is striking, though, is that near the fermionization limit (see
$g=25$) again a simple picture emerges: A fast, larger-amplitude
motion is superimposed on a slightly slower tunneling oscillation
whose period roughly equals that of the Rabi oscillations.

\begin{figure}
\includegraphics[width=0.85\columnwidth,keepaspectratio]{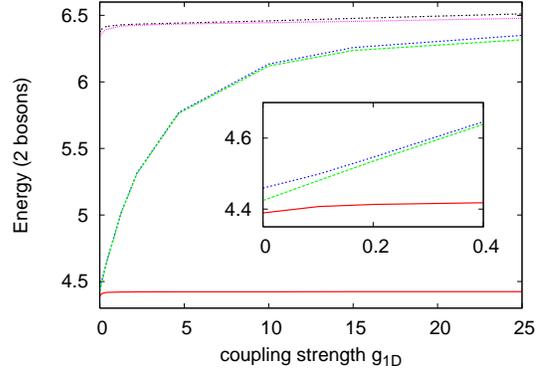}

\caption{(color online) Low-lying spectrum of two bosons in a double well
as a function of the interaction strength $g$. \emph{Inset}: Doublet
formation with increasing $g$.\label{cap:spec-h8}}
\end{figure}
To get an understanding of the oscillations, Fig.~\ref{cap:spec-h8}
explores the evolution of the two-body spectrum $\{ E_{m}(g)\}$ as
$g$ is varied. In the noninteracting case, the low-lying spectrum
is given by distributing the $N$ atoms over the lowest anti-/symmetric
orbital of the trap. This yields the $N+1$ energies $\{ E_{m}=E_{0}+m\Delta^{(0)}\}_{m=0}^{N}$,
where $\Delta^{(0)}=\epsilon_{1}-\epsilon_{0}$ is the energy gap
between these two orbitals, or the splitting of the lowest \emph{band}.
Assuming that for sufficiently small $g$ still only $N+1=3$ levels
are populated, then the imbalance $\delta(t)$ (and likewise $p_{\mathrm{R}}$)
can be computed to be \cite{zoellner07b} \begin{equation}
\delta(t)=\delta_{01}\cos(\omega_{01}t)+\delta_{12}\cos(\omega_{12}t),\label{eq:pop-imb}\end{equation}
where $\omega_{mn}=E_{m}-E_{n}$ and $\delta_{mn}=4\langle\Psi_{m}|\Theta(x)|\Psi_{n}\rangle c_{m}c_{n}$
are determined by the participating many-body eigenstates and their
weight coefficients $c_{m}$. At $g=0$, due to the levels' equidistance,
only a single mode with Rabi frequency $\omega_{01}=\omega_{12}=\Delta^{(0)}$
contributes. However, as the interaction is {}``switched on'', the
two upper lines $E_{1,2}$ virtually glue to one another to form a
doublet, whereas the gap to $E_{0}$ increases (Fig.~\ref{cap:spec-h8},
inset). For times $t\ll T_{12}\equiv2\pi/\omega_{12}$, we only see
an oscillation with period $T_{01}\ll T_{12}$, offset by $\delta_{12}$,
which on a longer timescale modulates the slower oscillation determined
by $\omega_{12}$. For small initial imbalances, $\left|c_{0}/c_{2}\right|=\left|\delta_{01}/\delta_{12}\right|\gg1$;
so for short times we observe the few-body analogue of Josephson tunneling.
In our case of an almost complete imbalance, in turn, $|\delta_{12}|$
dominates, which ultimately corresponds to \emph{self-trapping}, viz.,
extremely long tunneling times. These considerations convey a simple
yet essentially exact picture for the two-body counterpart of the
crossover from Rabi oscillations to self-trapping beyond the bare
two-mode approach common for condensates \cite{milburn97}.

It is obvious that the two-frequency description above breaks down
as the gap to higher-lying states melts, as for $g=4.7$.  Concordantly,
the dynamics becomes more complicated. However, in the fermionization
limit (exemplified for $g=25$), the system becomes integrable again
by mapping it to noninteracting fermions \cite{girardeau60}. As an
idealization, assume that at $t=0$ we put two (auxiliary) fermions
in the ground state of the \emph{right} well, where they would occupy
the lowest two orbitals. Expressing this through the fermionic eigenstates
$|\boldsymbol{n}\rangle_{\!-}$ of the full system leads to \cite{zoellner07b}
$\Psi(t=0)=\frac{1}{2}\sum_{a,b\in\{0,1\}}|1_{a}^{(0)},1_{b}^{(1)}\rangle_{\!-}$,
where $1_{a}^{(\beta)}$ denotes occupation of the symmetric ($a=0$)
or antisymmetric ($a=1$) orbital in band $\beta$. Analyzing the
corresponding energies, one finds that the frequencies contributing
to the imbalance dynamics are exactly $\Delta^{(0)}$ (the lowest-band
Rabi frequency, corresponding to the longer tunneling period) and
$\Delta^{(1)}$ (the splitting of the upper band). This intriguing
result states that only \emph{two} modes determine the imbalance dynamics,
so that the strongly repulsive atoms coherently tunnel back and forth
almost like a single particle. As an illustration, snapshots of the
density at different $t$ are displayed in Fig.~\ref{cap:pop-N2}(b).
This demonstrates the tunneling of a \emph{fragmented} pair.

\begin{figure}
\includegraphics[width=0.8\columnwidth,keepaspectratio]{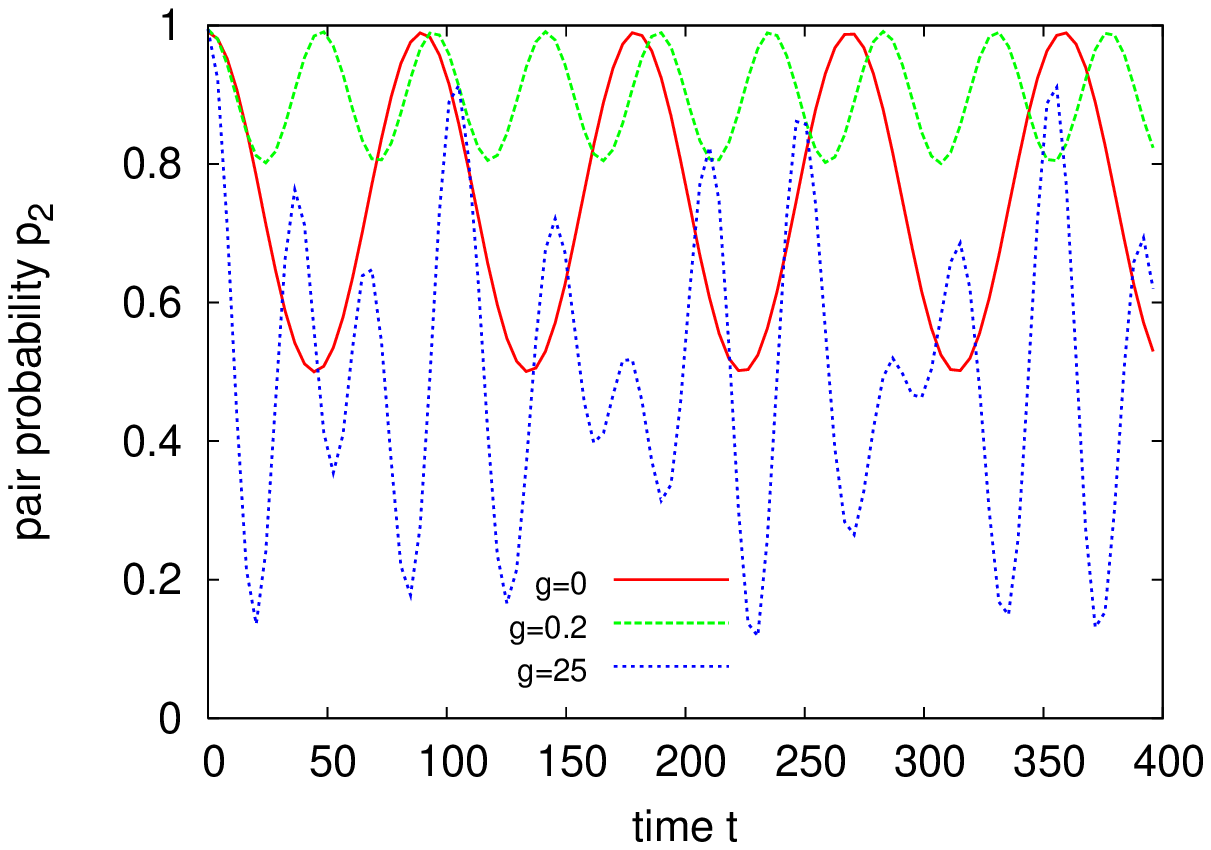}

\includegraphics[width=0.33\columnwidth,keepaspectratio]{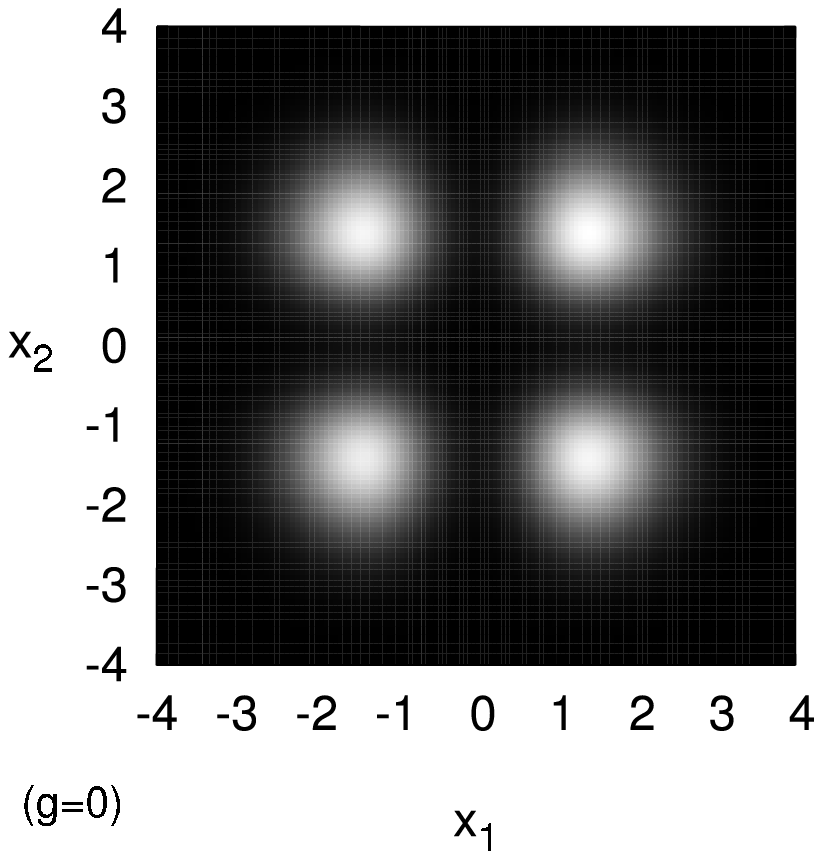}\includegraphics[width=0.33\columnwidth,keepaspectratio]{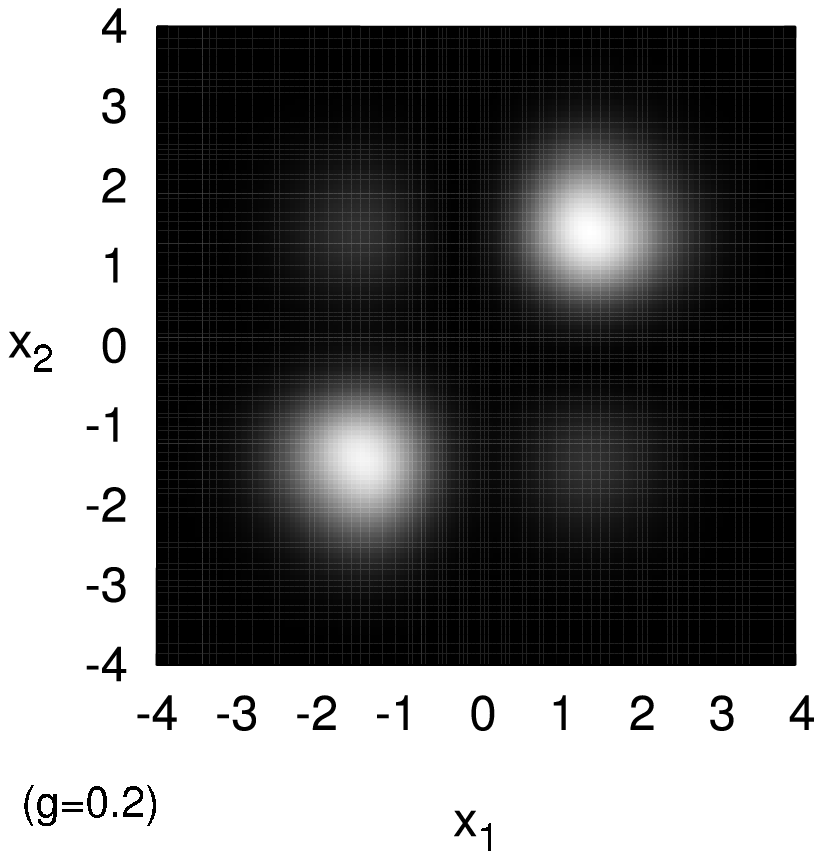}\includegraphics[width=0.33\columnwidth,keepaspectratio]{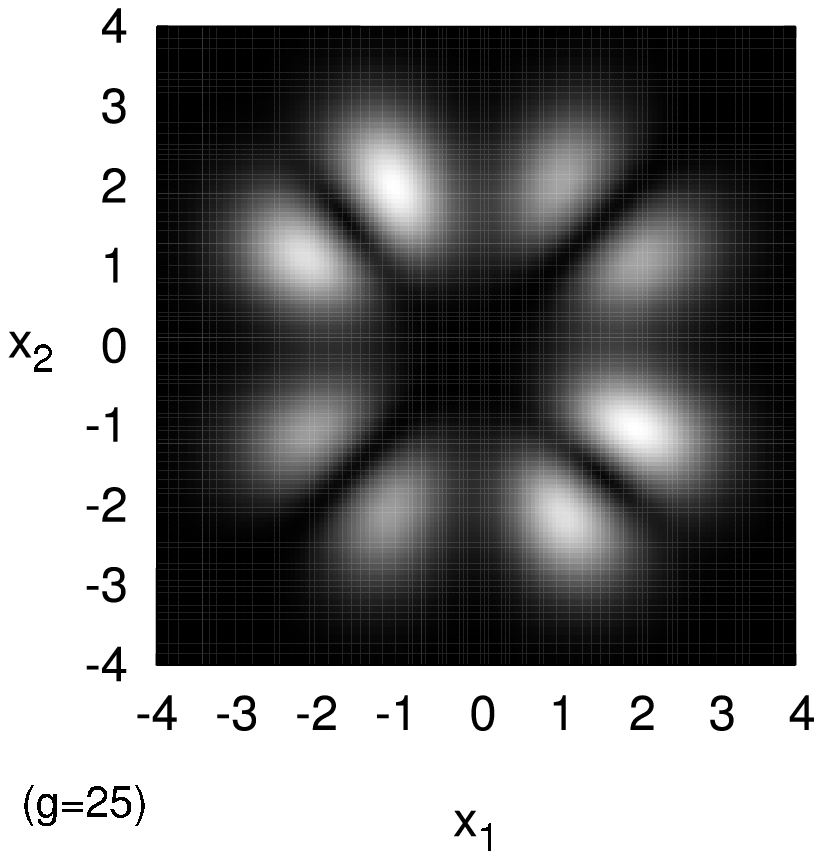}

\caption{(color online) \emph{Top}: Probability $p_{2}(t)$ of finding two
atoms in the same well for $g=0,0.2,25$. \emph{Bottom}: Snapshots
of two-body correlation function $\rho_{2}(x_{1},x_{2})$ at equilibrium
points, $\delta(t)=0$, for $g=0$ ($t=44$), $g=0.2$ ($t=128$),
and $g=25$ ($t=53$) -- \emph{from left to right}. \label{cap:pair}}
\end{figure}
In order to unveil the physical content behind the tunneling dynamics,
let us now investigate the two-body correlations. Noninteracting bosons
simply tunnel independently, which is reflected in the two-body density
(or correlation function) $\rho_{2}(x_{1},x_{2})$. As a consequence,
if both atoms start out in one well, then in the equilibrium point
of the oscillation it will be as likely to find both atoms in the
same well as in opposite ones. This is illustrated in Fig.~\ref{cap:pair},
which exposes $\rho_{2}$ at the equilibrium points and visualizes
the temporal evolution of the \emph{pair} (or \emph{same-site}) \emph{probability}
$p_{2}=\int_{\{ x_{1}\cdot x_{2}\ge0\}}\rho_{2}(x_{1},x_{2})dx_{1}dx_{2}$.
As we introduce small correlations, the pair probability does not
drop to $0.5$ anymore -- at $g=0.2$ it notably oscillates about
a value near 100\%. This is apparent from the equilibrium-point snapshot
of $\rho_{2}$: Both atoms remain essentially in the same well in
the course of tunneling. In other words, \emph{}they tunnel \emph{as
pairs}. On top of this, Fig.~\ref{cap:pair} in hindsight also lays
bare the nature of the fast (small-amplitude) modulations of $p_{\mathrm{R}}(t)$
encountered in Fig.~\ref{cap:pop-N2}(a) by linking them to temporary
reductions of the pair number $p_{2}$. Thus it is fair to interpret
them as attempted one-body tunneling. As before, the time evolution
becomes more involved as the interaction energy is raised to the fermionization
limit (cf. $g=25$). The two-body correlation pattern is fully fragmented
not only when the pair is captured in one well (corresponding, e.g.,
to the upper right corner $x_{1},x_{2}\ge0$), but also when passing
through the equilibrium point $t=53$. Similarly, the evolution of
$p_{2}(t)$ is governed by two modes, $\Delta^{(0)}\pm\Delta^{(1)}$,
and over time $p_{2}$ passes through just about any value from $1$
(fragmented pair) to almost zero (complete isolation).

\paragraph*{Many-body effects}

Although having focused so far on the case of $N=2$ atoms, the question
of higher atom numbers is interesting from two perspectives. For one
thing, it is fascinating because for $g\gg1$ many results become
explicitly $N$-dependent, including distinctions between even/odd
atom numbers \cite{zoellner06a}. (The experimental preparation of
definite $N=3,4,\dots$ is feasible, if harder to achieve due to losses.
In fact, the experimental setup in \cite{kinoshita04} requires only
an additional central barrier created by a Gaussian light sheet.)
On the other hand, in a setup consisting of a whole \emph{array} of
1D traps like in \cite{paredes04,kinoshita04}, number fluctuations
may automatically admix states with $N>2$.

\begin{figure}
\includegraphics[width=0.85\columnwidth,keepaspectratio]{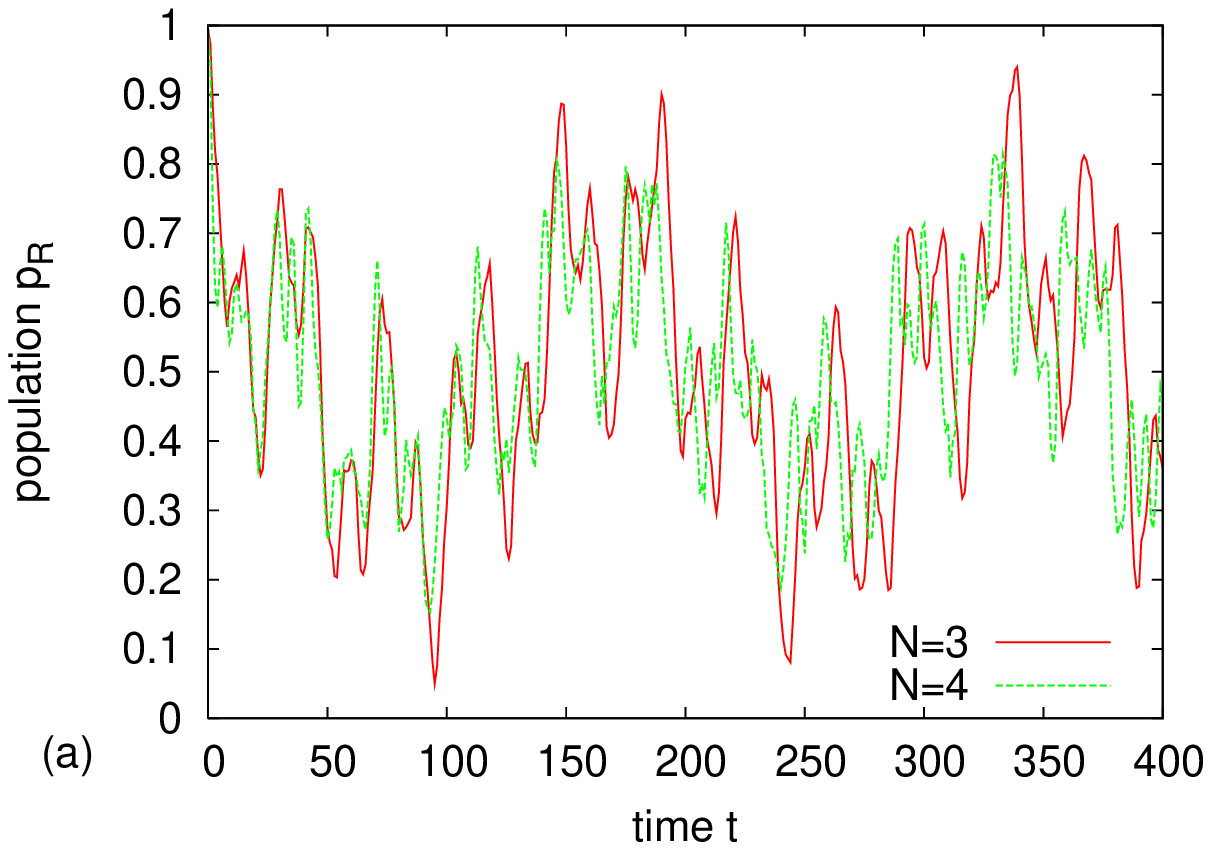}

\includegraphics[width=0.49\columnwidth,keepaspectratio]{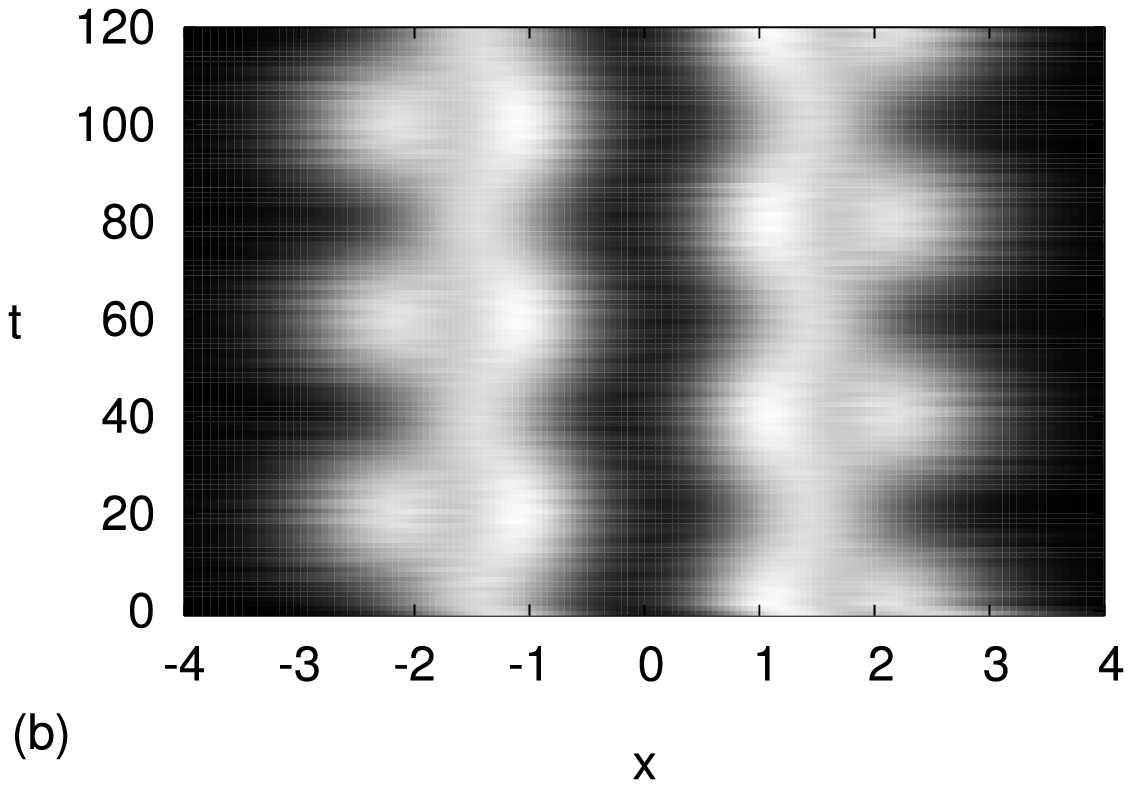}\includegraphics[width=0.49\columnwidth,keepaspectratio]{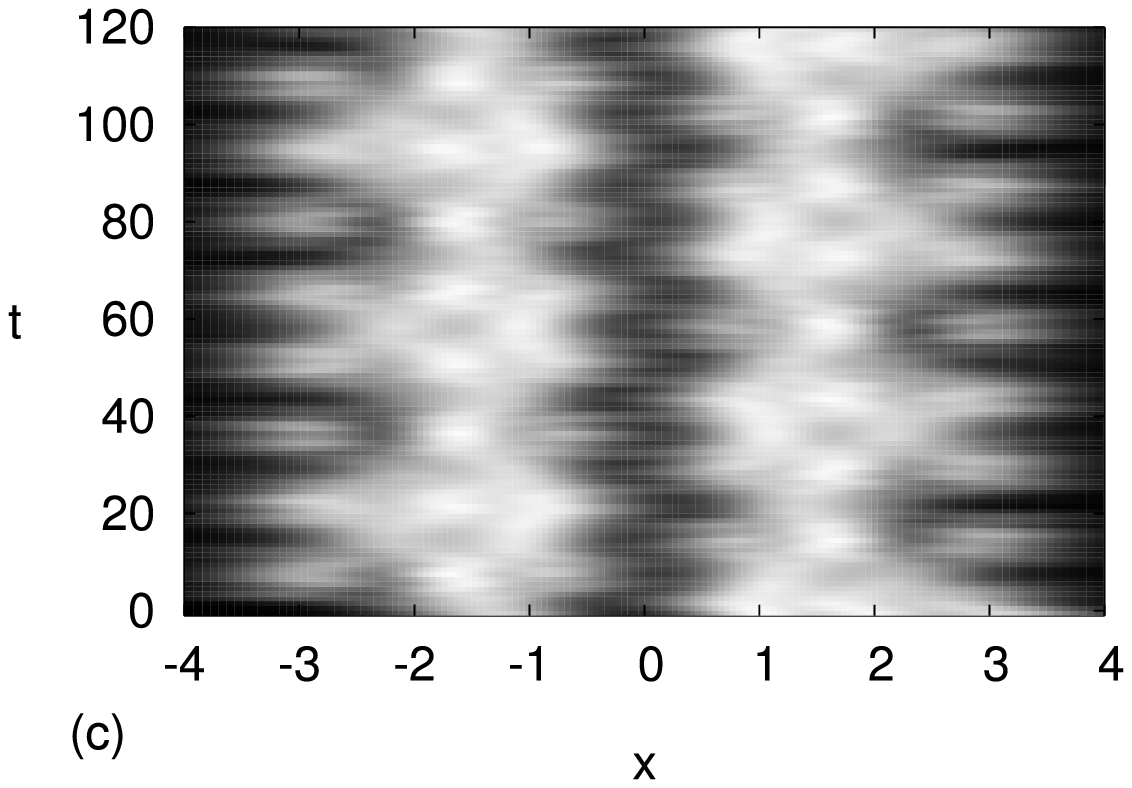}

\caption{(color online) Many-body effects in the fermionization limit ($g=25$).
\emph{}(a) Population of the right-hand well, $p_{\mathrm{R}}(t)$,
for $N=3,4$ atoms initially in one well. \emph{Bottom}: Density evolution
$\rho(x;t)$ for $N-1=2$ (b) and $N-1=3$ atoms (c) initially in
the right-hand well if exactly one atom is present on the left. \label{cap:N3-4}}
\end{figure}

For $N\ge3$, the weak-interaction behavior does not differ conceptually.
In fact, Eq. (\ref{eq:pop-imb}) carries over but with the sum now
running over $m<n\le N$. While the dynamics is no longer determined
by strictly two frequencies, the separation of time scales (related
to the formation of doublets in the spectrum) persists -- ultimately,
this should connect to the condensate dynamics valid for $N\gg1$.
Things become more intricate if we leave the two-mode regime, though.
In particular, the fermionization limit reveals a clear $N$ dependence
(Fig.~\ref{cap:N3-4}). Generally, an idealized state with $N$ fermions
initially in one well has contributions from all excitations $|1_{a_{0}}^{(0)},\dots,1_{a_{N-1}}^{(N-1)}\rangle_{\!-}$
($a_{\beta}=0,1$ $\forall\beta$) in the $N$ lowest bands. Hence
all tunnel splittings $\Delta^{(\beta)}$ for each band are expected
to be present \cite{zoellner07b}.  Figure~\ref{cap:N3-4}(a) conveys
an impression of the complexity of the dynamics by exhibiting $p_{\mathrm{R}}(t)$
for $N=3,4$. This somewhat erratic pattern may wash out the clear
signature of the two-atom case upon averaging over an array. In an
experiment, it is therefore desirable to reduce number fluctuations,
e.g., by having sufficiently high barriers in between different copies
of the double well\textbf{.}

In the context of many-body effects, it is interesting to consider
what happens if \emph{not} all $N\ge3$ atoms are prepared in one
well, but rather, say, $N-1$ in one well and one in the other. Paraphrased
in the case $N=3$, this is the question of the fate of an atom pair
if the target site is already occupied by an atom. The striking answer,
as evidenced in Fig.~\ref{cap:N3-4}(b), is that the process can
be viewed as single-atom tunneling on the background of the symmetric
two-atom ground state. The tunneling frequency in the fermionization
limit is simply the tunnel splitting $\Delta^{(1)}\approx2\pi/40$.
This has the intuitive interpretation of a fermion which---lifted
to the band $\beta=1$---tunnels independently of the two lowest-band
fermions. From that point of view, it should come as no surprise that
adding another particle destroys that simple picture. In fact, Fig.~\ref{cap:N3-4}(c)
reveals that if we start with $N-1=3$ atoms on the right, then the
tunneling oscillations appear erratic at first glance, and a configuration
with three atoms per site becomes an elusive event. (E.g., at $t\approx22$,
three atoms are on the left site, whereas at $t\approx44,\,72$ three
atoms are on the right.) In the spirit of the Fermi map above, this
can be understood as superimposed tunneling of one atom in the first
excited band ($\Delta^{(1)}$) and another in the second band ($\Delta^{(2)}\approx2\pi/15$),
while the remaining zeroth-band fermions stay inactive.

Finally, we mention that one may not only use the tilt $d$ to load
the atoms into one well, but also to study tunneling oscillations
in \emph{asymmetric} wells in order to actively tune the tunneling.
 A detailed investigation \cite{zoellner07b} reveals that, for
medium $g$, single-particle tunneling can be resonantly \emph{enhanced}
if the right well is lowered enough to compensate the interaction-energy
shift. In the fermionization limit, in turn, single-atom tunneling
turns out resonant already for $d=0$, while tuning $d$ makes other
resonances accessible.

In conclusion, we have performed an \emph{ab initio} investigation
of the full crossover from uncorrelated to fermionized tunneling of
a boson pair in a double well. Remarkable features of this pathway
are the strongly delayed pair tunneling encountered for medium interactions
and, in the fermionization limit, fragmented pair tunneling at the
Rabi frequency. Having pushed the notion of tunneling toward strongly
interacting systems, this opens up intriguing perspectives, ranging
from resonantly tuning the tunneling to considering multi-well setups.

Financial support from the Landesstiftung Baden-Württemberg is acknowledged
by P.S. and S.Z. The authors thank S.~Jochim and O.~Alon for discussions.\bibliographystyle{prsty}
\bibliography{/home/sascha/paper/pra/DW/phd,/home/sascha/bib/mctdh}

\end{document}